\documentclass[10pt,conference]{IEEEtran}
\IEEEoverridecommandlockouts

\AtBeginDocument{%
  \providecommand\BibTeX{{%
    \normalfont B\kern-0.5em{\scshape i\kern-0.25em b}\kern-0.8em\TeX}}}

\usepackage[ruled,vlined,lined,commentsnumbered]{algorithm2e}
\usepackage{booktabs}
\usepackage{balance}
\usepackage{moreverb}
\usepackage{fontenc}
\usepackage{amsmath}

\usepackage{amssymb}
\usepackage{fancybox}
\usepackage{color}
\usepackage{colortbl}
\usepackage{array}
\usepackage{diagbox}
\usepackage{flushend}
\usepackage{multirow}
\usepackage{multicol}
\usepackage{makecell}
\usepackage{graphicx}
\usepackage{setspace}
\usepackage{soul}
\usepackage[utf8]{inputenc}
\usepackage{fontawesome}
\usepackage{dialogue}
\usepackage{listings}
\usepackage{csvsimple}
\usepackage{longtable}
\usepackage[skip=0pt,labelfont=bf]{caption}
\usepackage[breaklinks]{hyperref}
\usepackage{url}
\usepackage{lscape}
\usepackage{rotating}
\usepackage{tikz}
\usepackage{enumitem}
\usepackage{wrapfig}
\usepackage[most]{tcolorbox}
\makeatletter
\@namedef{ver@lineno.sty}{9999/12/31}
\@namedef{opt@lineno.sty}{}
\makeatother
\usepackage{threeparttable}

\usepackage{marvosym}
\usepackage{pifont}

\newcolumntype{L}[1]{>{\raggedright\let\newline\\\arraybackslash\hspace{0pt}}m{#1}}
\newcolumntype{C}[1]{>{\centering\let\newline\\\arraybackslash\hspace{0pt}}m{#1}}
\newcolumntype{R}[1]{>{\raggedleft\let\newline\\\arraybackslash\hspace{0pt}}m{#1}}

\definecolor{codegreen}{rgb}{0,0.6,0}
\definecolor{codegray}{rgb}{0.5,0.5,0.5}
\definecolor{codepurple}{rgb}{0.58,0,0.82}
\definecolor{backcolour}{rgb}{0.95,0.95,0.92}
\definecolor{lightgreen}{HTML}{99d8c9}
\definecolor{darkred}{RGB}{192,0,0}
\definecolor{darkgreen}{RGB}{84,130,53}

\lstdefinestyle{mystyle}{
    commentstyle=\color{codegreen},
    keywordstyle=\color{magenta},
    numberstyle=\tiny\color{black},
    stringstyle=\color{codepurple},
    basicstyle=\footnotesize,
    breakatwhitespace=false,
    breaklines=true,
    captionpos=b,
    keepspaces=true,
    showspaces=false,
    showstringspaces=false,
    showtabs=false,
    tabsize=2
}

\setlist{noitemsep} 

\definecolor{darkpastelred}{rgb}{0.76, 0.23, 0.13}
\definecolor{ao(english)}{rgb}{0.0, 0.5, 0.0}

\definecolor{darkpastelred}{rgb}{0.76, 0.23, 0.13}
\definecolor{ao(english)}{rgb}{0.0, 0.5, 0.0}
\lstdefinelanguage{diff}{
	morecomment=[f][\color{blue}]{@@},     
	morecomment=[f][\color{red}]-,         
	morecomment=[f][\color{codegreen}]+,       
	morecomment=[f][\color{red}]{---}, 
	morecomment=[f][\color{codegreen}]{+++},
}

\definecolor{yellow}{RGB}{255,255,153}
\definecolor{grey}{RGB}{224,224,224}

\newboolean{showcomments}
\setboolean{showcomments}{true}
\ifthenelse{\boolean{showcomments}}
 { \newcommand{\mynote}[2]{
      \fbox{\bfseries\sffamily\scriptsize#1}
        {\small$\blacktriangleright$\textsf{\emph{#2}}$\blacktriangleleft$}}}
        { \newcommand{\mynote}[2]{}}

\definecolor{DarkOrange}{rgb}{0.8,0.3,0.0}
\definecolor{DarkCyan}{rgb}{0.0, 0.55, 0.55}

\newcolumntype{?}{!{\vrule width 1pt}}

\definecolor{grey}{rgb}{0.9,0.9,0.9}
\definecolor{lightgrey}{HTML}{f0f0f0}
\definecolor{mygreen}{HTML}{02818a}
\definecolor{mygray}{HTML}{666666}

\newcommand{\llmao}{LLMAO\xspace}
\newcommand{\grace}{GRACE\xspace}
\newcommand{\llmaoOchiai}{LLMAO$_{Ochiai}$\xspace}
\newcommand{\chatgptOchiai}{ChatGPT$_{Ochiai}$\xspace}
\newcommand{\toolname}{{\sc AgentFL}\xspace}
\newcommand{\bftoolname}{\textbf{\textsc{AgentFL}}\xspace}
\newcommand*{\eg}{e.g., }

\newcommand*{\ie}{i.e., }

\newcommand{\etal}{\emph{et~al.}\xspace}

\newcommand{\find}[1]{
\begin{tcolorbox}[tile,fontupper=\small,size=fbox,boxsep=2.2mm,boxrule=0pt,top=0pt,bottom=0pt,borderline west={0.5mm}{0pt}{black!50!white},colback=black!5!white]
#1
\end{tcolorbox}
}

\newcommand{\notez}[1]{
\begin{tcolorbox}[size=fbox,boxrule=0.5pt,top=0.5pt,bottom=0.5pt,
colframe=blue!5!black,colback=black!5!white]
\em #1
\end{tcolorbox}
}

\begin{document}

\title{\toolname: Scaling LLM-based Fault Localization to Project-Level Context}
\author{
    \IEEEauthorblockN{
        Yihao Qin\IEEEauthorrefmark{1},
        Shangwen Wang\IEEEauthorrefmark{1},
        Yiling Lou\IEEEauthorrefmark{2},
        Jinhao Dong\IEEEauthorrefmark{3},
        Kaixin Wang\IEEEauthorrefmark{2},
        Xiaoling Li\IEEEauthorrefmark{1},
        Xiaoguang Mao\IEEEauthorrefmark{1},
    }
    \IEEEauthorblockA{\IEEEauthorrefmark{1}National University of Defense Technology, China,\\
    \{yihaoqin, wangshangwen13, lixiaoling, xgmao\}@nudt.edu.cn
    }
    \IEEEauthorblockA{\IEEEauthorrefmark{2}Fudan University, China, \{yilinglou@, kxwang23@m.\}fudan.edu.cn
    }
    \IEEEauthorblockA{\IEEEauthorrefmark{3}Peking University, China, dongjinhao@stu.pku.edu.cn
    }
    
    \thanks{* Our work has been accepted by Transactions on Software Engineering (TSE) in 2025. For the latest version, please refer to ``SOAPFL: A Standard Operating Procedure for LLM-based Method-Level Fault Localization''.}
}

\maketitle

\begin{abstract}
Fault Localization (FL) is an essential step during the debugging process. With the strong capabilities of code comprehension, the recent Large Language Models (LLMs) have demonstrated promising performance in diagnosing bugs in the code. 
Nevertheless, due to LLMs' limited performance in handling long contexts, existing LLM-based fault localization remains on localizing bugs within a \textit{small code scope} (i.e., a method or a class), which struggles to diagnose bugs for a \textit{large code scope} (i.e., an entire software system). 
To address the limitation, this paper presents \bftoolname, a multi-agent system based on ChatGPT for automated fault localization.
By simulating the behavior of a human developer, \bftoolname models the FL task as a three-step process, which involves comprehension, navigation, and confirmation.
Within each step, \bftoolname hires agents with diversified expertise, each of which utilizes different tools to handle specific tasks.
Particularly, we adopt a series of auxiliary strategies such as Test Behavior Tracking, Document-Guided Search, and Multi-Round Dialogue to overcome the challenges in each step.
The evaluation on the widely used Defects4J-V1.2.0 benchmark shows that \bftoolname can localize 157 out of 395 bugs within Top-1, which outperforms the other LLM-based approaches and exhibits complementarity to the state-of-the-art learning-based techniques.
Additionally, we confirm the indispensability of the components in \bftoolname with the ablation study and demonstrate the usability of \bftoolname through a user study.
Finally, the cost analysis shows that \bftoolname spends an average of only 0.074 dollars and 97 seconds for a single bug.

\end{abstract}


\section{Introduction}
\label{sec:intro}
Fault Localization (FL) is an important but time-consuming phase in the software debugging process. In particular, developers can spend nearly half of the debugging time to understand and localize the bug before they fix the buggy code locations~\cite{alaboudi2021exploratory}.
To release developers from the laborious FL tasks, various techniques have been proposed to automatically localize buggy program entities (e.g., classes, methods, or statements), among which the spectrum-based fault localization (SBFL)~\cite{abreu2009sbfl,raselimo2019sbfl,Reis2019sbfl,wong2014dstar,zhang2011sbfl} and learning-based fault localization (LBFL)~\cite{zheng2016dnnfl,sohn2017fluccs,zhang2019cnnfl,li2019deepfl} have been extensively studied and perform progressive effectiveness.
As one of the most famous FL techniques, SBFL statistically analyzes the coverage information and prioritizes the program entities covered by more failed tests and fewer passed tests.
To better utilize the coverage information, LBFL techniques such as \grace~\cite{lou2021grace} represent coverage with a graph structure and leverage GNN models to learn useful features. Several state-of-the-art techniques have also suggested incorporating auxiliary information, such as code complexity and code history, and leveraging the combined features of such information with machine/deep learning models~\cite{sohn2017fluccs,li2019deepfl}.

The recent advance of Large Language Models (LLMs) has shed new light on fault localization. Having been trained on massive code and textual data, LLMs exhibit a strong capability in code comprehension~\cite{pu2023summarization,li2023explaining, yuan2023evaluating}, including detecting and localizing bugs in the code~\cite{wu2023llmfl}.
Existing LLM-based fault localization mainly remains on the preliminary application of LLMs via basic prompting or fine-tuning. In particular, Wu \etal~\cite{wu2023llmfl} prompt ChatGPT with simple instructions (e.g., {\em ``Please analyze the following code snippet for potential bugs...''}) and test failure information to localize bugs in the given method; \llmao~\cite{yang2024llmao} fine-tunes LLMs with adapters on the fault localization dataset (i.e., the tuning input are 128 consecutive code lines while the output are the buggy line labels). 
Although showing promising effectiveness, existing techniques can only address a simplified fault localization scenario, \ie localizing the buggy statements in the given buggy method.
The potential of LLMs under the project-level fault localization, \ie {\bf localizing the buggy methods from an entire project}, has not been thoroughly studied, possibly due to the following technical challenges.
First, the codebase of a real-world project often contains millions of tokens, which are far beyond the maximum input length of existing LLMs (\eg the input lengths of the GPT-3.5 series models vary from 4,096 to 16,385 tokens). Therefore, it is often infeasible and unacceptably resource-consuming to directly query LLMs with the entire project for fault localization. 
Second, existing LLMs exhibit deteriorating performance as the length of input context increases and have a tendency to overlook critical information within the long inputs~\cite{liu2024lost}. Therefore, simply feeding LLMs with as much code as possible might not be an optimal way to leverage the superior capabilities of the model.


To scale up LLM-based fault localization to the project-level contexts, in this work, we propose to decompose the localization process into multiple phases and selectively enhance LLMs with different debugging information during each phase. We are inspired by the debugging process of human developers, who often take various actions such as {\em analyzing the fault} (comprehending an error trace or a core dump), {\em browsing the codebase} (following each computational step in the execution of the failing test), and {\em examining the code} (building mental representation by reading and understanding the code)~\cite{alaboudi2021exploratory,bohme2017bug}.
Specifically, by decomposing the localization process into several steps and achieving FL through the collaboration of multiple LLM-driven agents (\ie intelligent entities that can perceive the environment, make decisions, and perform actions), the advantages we expect are twofold.
On the one hand, this multi-step approach allows more controllable model inputs, which avoids excessively long contexts by merely providing essential information for the LLMs in each step. 
On the other hand, the capabilities of LLMs are further exploited since each agent focuses on processing specific tasks with their customized expertise.

Based on the above intuition, we propose a novel LLM-based FL system, \toolname, which incorporates multiple LLM-driven agents to localize bugs for an entire project.
To mitigate the inherent limitations of LLMs in handling large-scale codebases, \toolname decomposes the project-level fault localization into three stages, including {\bf Fault Comprehension}, {\bf Codebase Navigation}, and {\bf Fault Confirmation}, where each stage is driven by multiple specialized agents. 
(1) The fault comprehension stage leverages LLMs to analyze the potential causes of the exposed fault, which would serve as the essential guidelines for the following two stages. Different from existing LLM-based FL techniques that only provide LLMs with the error message and the failed test code, \toolname incorporates a novel {\bf Test Behavior Tracking} approach to collect the complete test execution behaviors (\ie the test utility code) via lightweight program instrumentation. 
(2) The codebase navigation stage then leverages LLMs to identify suspicious methods in a gradual refinement way, which first narrows down the faulty code space by finding out the suspicious classes and then selects related methods from these classes. To achieve this, \toolname incorporates a novel {\bf Document-Guided Search} approach, which utilizes both existing documents and LLM-enhanced documents to identify classes and methods related to the potential error causes. 
(3) The fault confirmation stage lastly leverages LLM to revisit all the suspicious methods (identified in the previous stage) and decide the most suspicious one as the final result.  In this stage, \toolname incorporates a novel {\bf Multi-Round Dialogue} approach, which iteratively asks LLM to review each suspicious method with its detailed information.


We evaluate \toolname on the Defects4J-V1.2.0~\cite{just2014defects4j} benchmark which includes 395 real-world bugs from 6 Java projects.
To compare with the existing LLM-based FL techniques, we empower them to achieve project-level localization with Ochiai~\cite{abreu2006ochiai}.
The results show that \toolname can localize 157 bugs within Top-1, which significantly outperforms the LLM-based baselines \chatgptOchiai~\cite{wu2023llmfl} and \llmaoOchiai~\cite{yang2024llmao}.
\toolname also shows complementarity with existing learning-based techniques such as  DeepFL~\cite{li2019deepfl} and \grace~\cite{lou2021grace}.
We further validate the importance of different components in \toolname with an ablation study.
In addition, through a user study, we show the usability of \toolname in practice with its ability to provide both suspicious methods and rationale. 
Finally, the cost analysis reveals that \toolname requires only an average of 0.074 dollars and 97 seconds to localize a fault.

The main contributions of this paper are:
\begin{itemize}[leftmargin=*]
    \item We propose to decompose the project-level fault localization into three stages, \ie comprehension, navigation, and confirmation. Such a process adheres to the developers' debugging practice and holds the potential for effectively managing the inputs of the LLMs.
    \item We present \toolname, a three-stage FL technique empowered by multiple LLM-driven agents, capable of automatically localizing faults within the entire software.
    \item We extensively evaluate \toolname on both Defects4J-V1.2.0 and Defects4J-V2.0.0. \toolname outperforms the other LLM-based approaches and shows complementarity with existing LBFL techniques on both datasets. Additionally, we conduct the user study and cost analysis to demonstrate the usability of \toolname.
\end{itemize}

\section{Background \& Related work}
\label{sec:related_work}


\subsection{Statistic-Based Fault Localization}
Spectrum-based and learning-based fault localization are mainstream FL technologies that achieve prominent performance on the Defects4J~\cite{just2014defects4j} benchmark.
Given the buggy program and a set of tests including both passed and failed instances, the spectrum-based fault localization (SBFL) localizes suspicious program entities through the hypothesis that the fault location should be covered by more failed tests than passed ones.
To achieve this, SBFL first collects coverage information for each program element $e$ by recording the number of passed tests $T_{p}(e)$ and failed tests $T_{f}(e)$ that cover $e$, then designs different formula such as Ochiai~\cite{abreu2006ochiai}, DStar~\cite{wong2013dstar} and Jaccard~\cite{abreu2007jaccard} to calculate the suspiciousness scores for ranking all the elements.
As an example, Ochiai computes the suspiciousness of an element $e$ as $T_{f}(e)/((T_{f}(e)+T_{p}(e))T_{f})^{1/2}$, where $T_{f}$ is the number of all failed tests.

To use coverage more exhaustively and combine extra valuable information such as code history, learning-based fault localization (LBFL) is also extensively studied.
FLUCCS~\cite{sohn2017fluccs} boosts SBFL with code and change metrics.
DeepFL~\cite{li2019deepfl} learns to combine four dimensions of knowledge from SBFL, mutation-based FL, code complexity, and text similarity through RNN~\cite{tomas2010rnn} and MLP~\cite{popescu2009mlp}.
\grace~\cite{lou2021grace} integrates coverage with fine-grained code structures by a GNN~\cite{li2016ggnn} model.
Other well-known LBFL techniques include CNNFL~\cite{zhang2019cnnfl}, TraPT~\cite{li2017trapt}, CombineFL~\cite{zhou2021combinefl}, and GNet4FL~\cite{qian2023gnet4fl}, etc.

Based on the high dependence of SBFL and LBFL on coverage information and code features, we consider them both as statistic-based fault localization (STBFL) to distinguish them from the approaches based on the LLMs.

\subsection{LLM-Based Fault Localization}
With the emergence of the Large Language Model (LLM), LLMs such as ChatGPT~\cite{openai}, Phind~\cite{phind}, CodeGen~\cite{nijkamp2023codegen}, and StarCoder~\cite{li2023starcoder} have shown remarkable power on aligning semantics between natural languages (NL) and programming languages (PL).
In the ChatGPT-4 launch event, OpenAI prompted the model to localize and fix errors by providing program code and error log, which unveiled the potential and applicability of ChatGPT-4 in fault localization.

However, due to the well-known context limitation issue as well as the decreasing performance on longer input contexts~\cite{wu2023llmfl,liu2024lost}, existing LLM-based FL techniques primarily concentrate on localizing buggy lines within a code snippet, while neglecting the identification of bugs in an entire software project.
Given a context of 128 code lines, \llmao~\cite{yang2024llmao} trains lightweight bidirectional adapters on top of LLMs to produce a suspiciousness score for each line.
Wu \etal~\cite{wu2023llmfl} prompts ChatGPT with code and error logs, and instructs the model to identify the buggy lines. 
To assist developers from another perspective, in this paper, we endow the LLMs with the ability to automatically identify bugs in the entire software.

\subsection{LLM-driven agents \& SOP}
Supported by the existing studies about human debugging behaviors~\cite{alaboudi2021exploratory,bohme2017bug}, we regard fault localization as a complex task, which is difficult to get desired results with only one LLM inference.
Therefore, we inherited the design principles of LLM-driven agents and standard operating procedure (SOP) during the construction of \toolname.

LLM-driven agents are designed to solve complex tasks specified in natural language autonomously.
Equipped with mechanisms including memory management~\cite{zhou2023recurrentgpt}, tool usage~\cite{patil2023gorilla}, and multi-agent communication~\cite{li2023camel}, LLM-driven agents have been proposed to conduct various tasks such as social behavior simulation~\cite{park2023agent}, website tasks~\cite{gur2024realworld}, and interactive writing~\cite{zhou2023recurrentgpt}.
There are also LLM-driven agents such as AutoGPT~\cite{autogpt} and BabyAGI~\cite{babyagi} that aim at conducting arbitrary missions with user instructions.
More recently, multi-agent systems such as MetaGPT~\cite{hong2024metagpt} and ChatDev~\cite{qian2023chatdev} have started to move towards automated software development. 

SOP is a concept from the real-world engineering domain, for LLM-driven agents, SOP is a symbolic plan to make the agents more controllable~\cite{hong2024metagpt,zhou2023agents}.
For example, in MetaGPT, employed agents (e.g., product managers and engineers) work together by following a predefined and streamlined workflow, where all intermediate outputs (e.g., the document and code) are structured and standardized.
SOP plays a vital role in task decomposition and effective coordination, which ensures the robustness of the multi-agent system.

\section{Approach}
\label{sec:approach}
\toolname is a multi-agent fault localization system that localizes buggy methods for the given project in three steps. We first introduce an overview of \toolname and then explain the detailed steps respectively. 

\subsection{Overview}
\begin{figure}[!t]
    \centering
    \includegraphics[width=\linewidth]{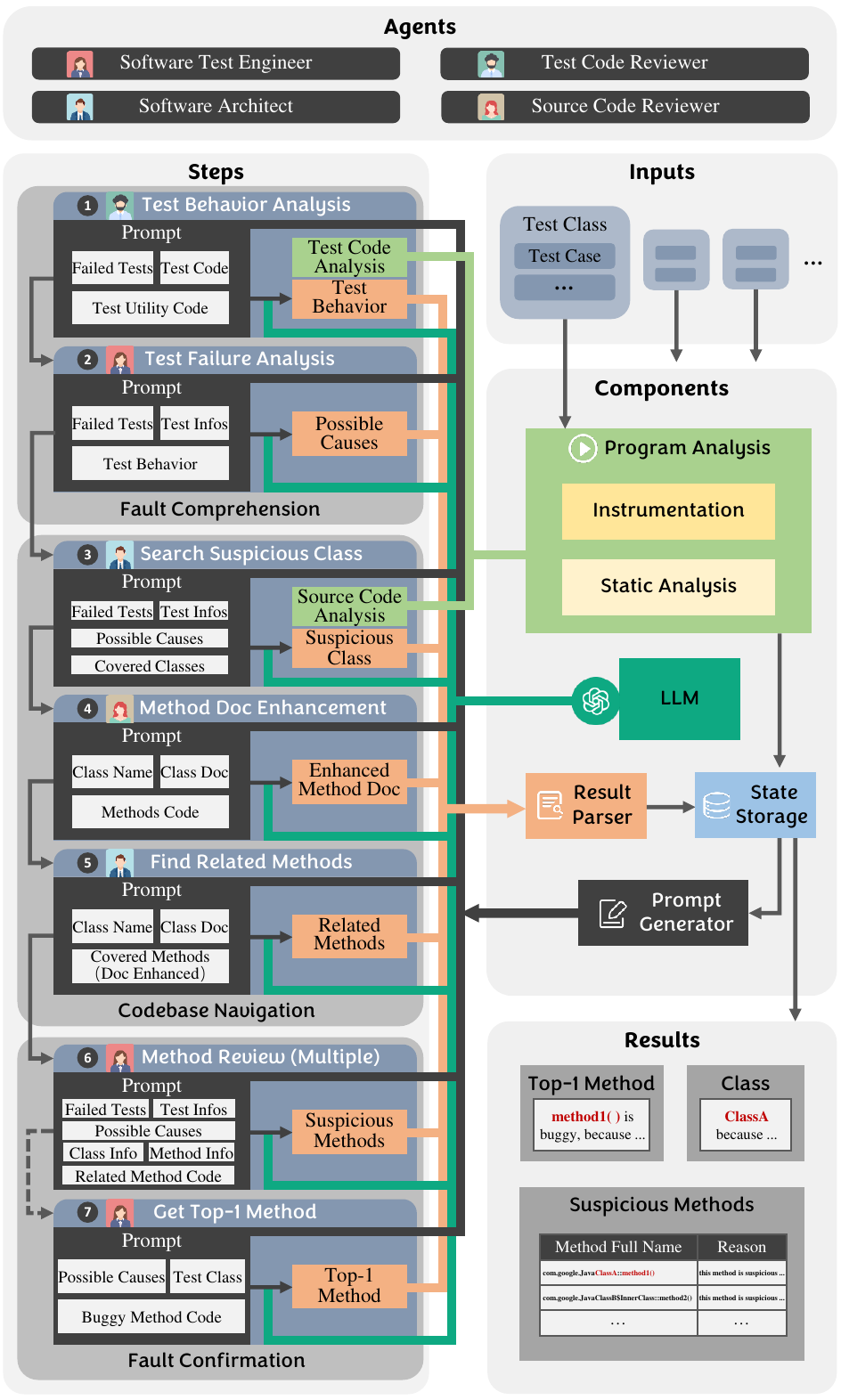}
    \caption{Overview of \toolname.}
    \label{fig:approach}
\end{figure}

The overall workflow of \toolname is shown in Figure \ref{fig:approach}. 
In particular, \toolname consists of three \textit{steps}; in each \textit{step}, \toolname is supposed to tackle different \textit{tasks}; to specialize the LLMs to solve different tasks, four different \textit{agents} are constructed by enhancing the LLMs with domain knowledge and external \textit{components}. 


{\bf Inputs \& Outputs.} The inputs of \toolname include the codebase of the entire project, the failed test methods (\ie the failed test cases), and the error message of the triggered fault. As it is common for one fault to trigger multiple failed test methods (\eg 22 test methods fail in the bug Chart-26 in Defects4J) and it is less efficient to diagnose each failed test method separately, \toolname focuses on one failed test class (\ie the aggregation of the failed test methods in the same class) in each run, and the total number of runs is the number of failed test classes. The outputs of \toolname include the buggy location and the corresponding explanations, which consist of the following three parts: (i) the {\em Top-1 Suspicious Method} along with its reason for being buggy, (ii) the suspicious {\em Class}, and (iii) other {\em Suspicious Methods} in the class along with their reasons for being buggy. 

{\bf Steps \& Tasks.}
By mimicking the human debugging strategies~\cite{alaboudi2021exploratory,bohme2017bug}, \toolname models the fault localization process as an SOP which comprises three steps, including (i) understanding the cause of the fault ({\em Fault Comprehension}), (ii) browsing the relevant program elements ({\em Codebase Navigation}), and (iii) validating the buggy location ({\em Fault Confirmation}). In each step, \toolname tackles different tasks (identified by \ding{182} - \ding{188} in Figure \ref{fig:approach}), \eg the first step of fault comprehension involves the \ding{182} \textit{Test Behavior Analysis} task and the \ding{183} \textit{Test Failure Analysis} task.

{\bf Agents.}
To specialize the LLMs to solve different tasks, \toolname incorporates four different LLM-driven \textit{agents} (\ie Test Code Reviewer, Source Code Reviewer, Software Architect, and Software Test Engineer), which enhance LLMs with task-specific knowledge and external capabilities (\eg program analysis component). In particular, the {\em Test Code Reviewer} agent (used in Task 1) is designed for reviewing the test code and summarizing the test behavior; the {\em Source Code Reviewer} agent (used in Task 4) is designed for writing high-quality code comments; the {\em Software Architect} agent  (used in Tasks 3\&5) is designed for software architecture and adept at finding areas of the program that may be problematic; the {\em Software Test Engineer} agent (used in Tasks 2\&6\&7) is designed for all the tasks associated with analyzing and validating the test failures. 
Each LLM-driven agent is customized by prompting LLMs with specific system instructions and enabling LLMs to use external components. In particular, the specific system instruction is typically provided at the start of the interaction to serve as the initial context or instruction for the LLMs.
Due to space constraints, we present the system instruction of the {\em Software Test Engineer} agent below:
\find{
You are a Software Test Engineer. We share a common interest in collaborating to successfully locate the buggy code that causes the test class to fail. Your main responsibilities include examining the information of the failed tests to analyze the possible causes of the test failures, and determining the method that needs to be fixed. To locate the bug, you must write a response that appropriately solves the requested instruction based on your expertise.
}

{\bf Components.}
The LLMs alone fall short in limited context length and unstable performance of hallucination~\cite{zhang2023siren}. Therefore, \toolname incorporates multiple components to enhance LLM-driven agents with memorization, reasoning, and tool usage capabilities. In other words, with the LLM as the brain of each agent, the other components provide LLMs with external capabilities of better utilizing debugging contexts. 
In particular, the {\em Program Analysis} component is driven by lightweight Instrumentation and Static Analysis, which equips the agents with the tools (Test/Source Code Analysis) to perceive the test execution process; the components of {\em Result Parser}, {\em State Storage}, and {\em Prompt Generator} enable a loop to drive the pipeline operations of multiple tasks. Specifically, the {\em Prompt Generator} collects material from the {\em State Storage} and generates the prompt according to a template predefined for each task.
After the agent returns the response, the {\em Result Parser} transforms the result into formatted data and sends it to the {\em State Storage}, where the result will be utilized in the downstream steps.
Moreover, the data in {\em State Storage} can also come from the {\em Program Analysis} component when the agent calls a tool.



\begin{figure}[!t]
    \centering
    \includegraphics[width=\linewidth]{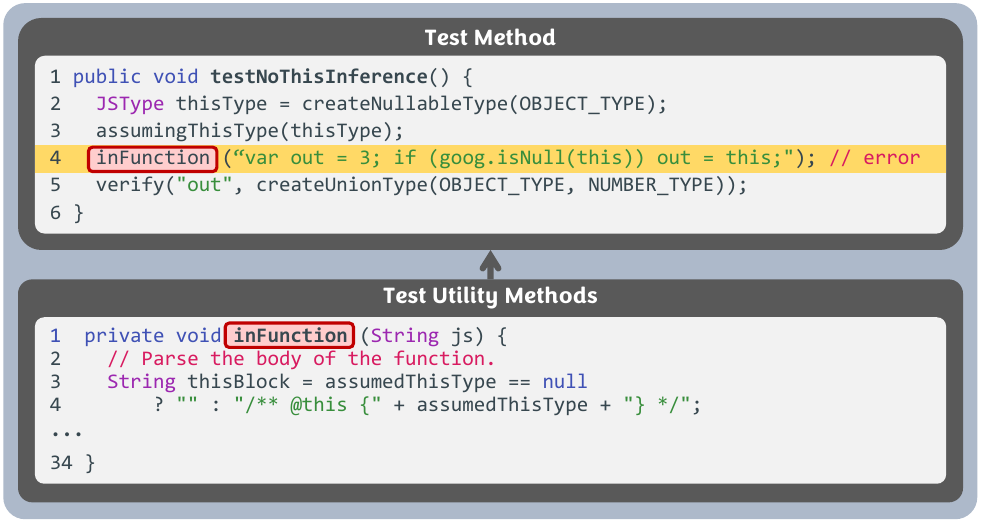}
    \caption{Example from Closure-19.}
    \label{fig:utility}
\end{figure}

\subsection{Fault Comprehension}
\label{subsec:compre}

{\bf Aim.}
To guide the whole localization process, \toolname starts with the {\em Fault Comprehension} step to better understand the cause of the fault before starting to find it.
Our intuition is that compared to direct reasoning, the ``understanding-reasoning'' manner bridges the test failure information and the buggy source code with intermediate thought, which is more akin to the human mindset.
Specifically, the {\em Fault Comprehension} step consists of two tasks.
Provided with the test code, the \ding{182} {\em Test Behavior Analysis} task prompts the LLM to describe in detail the behavior of the failed test cases.
In the \ding{183} {\em Test Failure Analysis} task, the test behavior we obtained before together with other test failure information (including error stack trace, test code, and test output) encourages the LLM to list the possible causes of the fault.
As the intermediate thought, the possible causes will inform the subsequent navigation step.

{\bf Challenge.}
Although the attendance of test code (i.e., the test method executed during each unit test) has been proved helpful in LLM-based FL~\cite{wu2023llmfl}, we observe that the LLMs may still suffer from the limited information in the test method.
Figure \ref{fig:utility} shows an example derived from bug Closure-19 in Defects4J-V1.2.0~\cite{just2014defects4j}, in this case, the test \texttt{testNoThisInference} failed when a test utility method \texttt{inFunction()} is called at line 4.
However, since the developers kept all the detailed test logic in test utility methods that are called by the test method, the LLMs may find it difficult to comprehend the complete test behavior through only the test method.
In particular, test utility methods are very prevalent in practice, e.g., each test method calls 4.93 other test utility methods on average in the Defect4J-V1.2.0 benchmark. Therefore, the information in the utility methods can be potentially helpful for comprehensively comprehending the test failure.


\begin{figure}[!t]
    \centering
    \includegraphics[width=\linewidth]{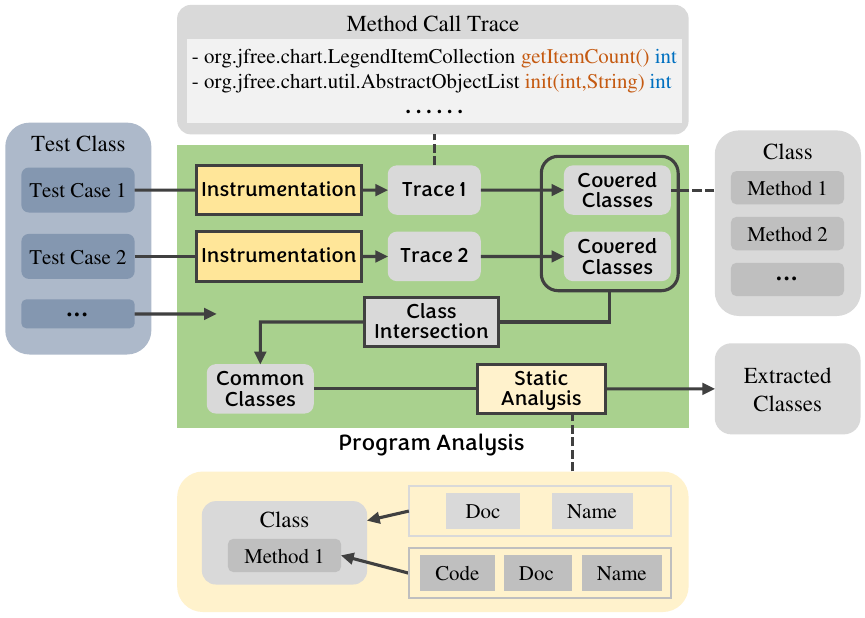}
    \caption{The process of the Program Analysis.}
    \label{fig:program_analysis}
\end{figure}

\begin{figure}[!t]
    \centering
    \includegraphics[width=\linewidth]{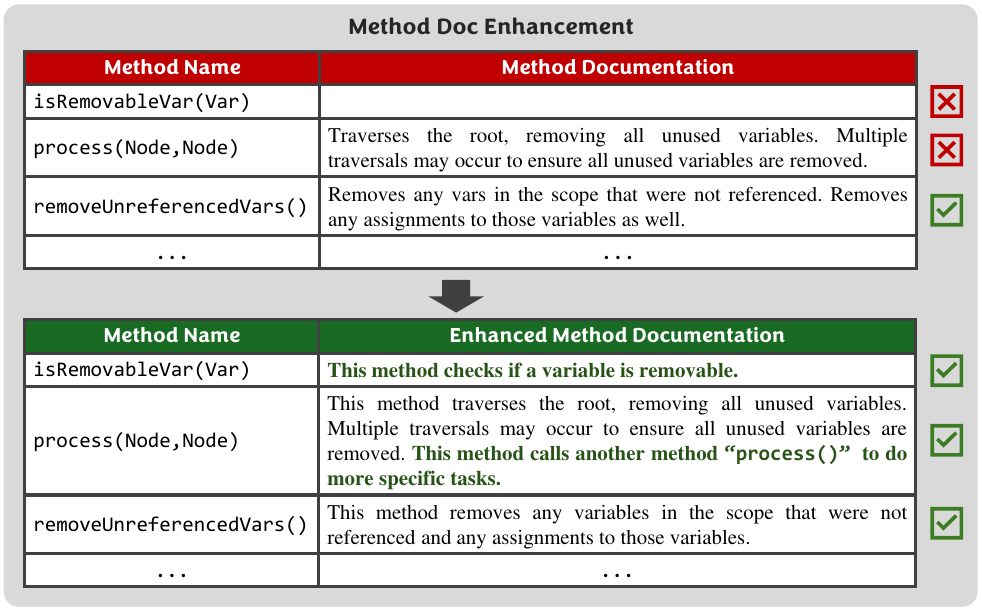}
    \caption{An example of method document enhancement.}
    \label{fig:doc_enhance}
\end{figure}

{\bf Strategy.}
To track the complete test execution behavior from the test utility methods, we adopt the {\bf Test Behavior Tracking} strategy which equips the LLMs with the capability of program analysis.
Specifically, the {\em Test Code Reviewer} agent in task \ding{182} utilizes the {\em Test Code Analysis} tool to fill the prompt with the {\em Test Utility Code} before it is asked to summarize the test behavior.
The more detailed process of Test Code Analysis in the {\em Program Analysis} component is demonstrated in Figure \ref{fig:program_analysis}.
Firstly, the tool applies lightweight {\em Instrumentation} for each failed test case to record the {\em Method Call Trace} during test code execution, and parses the trace to register the {\em Covered Classes}, each {\em Class} comprised of the full class name (e.g., \texttt{com.google.jscomp.TypeInferenceTest}) and a signature list (e.g., \texttt{init(int,String) int}) of the covered test utility methods in that class.
Secondly, a {\em Class Intersection} operation is conducted to keep the common classes and methods that are covered by all failed test cases, we refer to the output as {\em Common Classes}.
Finally, the {\em Static Analysis} further augments the classes and methods with the document, name, and code from the codebase.
As the output of the tool, {\em Extracted Classes} contains information about both the covered test class and the covered test utility methods, which is then reserved in the {\em State Storage} and used to construct the {\em Test Utility Code} of the prompt in {\em Test Behavior Analysis} task.


\subsection{Codebase Navigation}
\label{subsec:navi}

{\bf Aim.}
Provided with the comprehension of the possible causes in the last step, the {\em Codebase Navigation} step intends to identify all related methods from the entire codebase.
Our intuition is that given the test failure information and the possible causes generated from task \ding{183}, ChatGPT can gradually browse the entire codebase from global to local and identify the program entities that may be responsible for the test failure.
Specifically, we start with two tasks for the {\em Codebase Navigation} step.
The \ding{184} {\em Search Suspicious Class} task intends to localize the most suspicious class among all covered classes during a test execution process.
After that, the \ding{186} {\em Find Related Methods} task aims to filter out all the methods that may relate to the bug within the suspicious class.
To depict the role of methods in a class, the information about the suspicious class is also included in the prompt.

{\bf Challenge.}
Nevertheless, the above design still faces three challenges: (1) Due to the possible invalidity or weakness of LLMs under long contexts~\cite{liu2024lost}, a strategy should be taken for leading the model to concentrate on critical information during the browsing process;
(2) An excessive number of covered classes can still result in LLM distraction and context length issues, for example, the failed integration test \texttt{testSingletonGetter1} of bug Closure-36 covers hundreds of methods in 208 classes;
(3) The ChatGPT model struggles to identify related methods directly with the existing documents due to the absence of some method comments and the ``function nested'' problem.
The upper side of Figure \ref{fig:doc_enhance} shows an example of the original method comments in bug Closure-1.
In the first row, the absence of the comment of method \texttt{isRemovableVar(Var)} prevents ChatGPT from better understanding its functionality.
In the second row, although the comment of \texttt{process(Node,Node)} says it ``{\em removes all unused variables}'', this method accomplishes the functionality by calling other methods.
We refer to this inconsistency of comment and function of a method as the ``function nested'' problem, which could mislead the LLMs to focus on irrelevant methods.

{\bf Strategy.}
To alleviate the aforementioned challenges, we adopt a series of heuristic strategies:
(1) Considering that the LLMs can associate the function of a class/method with the function that may cause the bug through the proximity of natural language description, we adopt a {\bf Document-Guided Search} strategy to first search for suspicious classes and then find the related methods.
Specifically, in task \ding{184}, we extract the covered classes by source code analysis (Figure \ref{fig:program_analysis}), and prompt ChatGPT to select a single class from a markdown format table built with names and documentation of the covered classes.
After localizing the suspicious class, the \ding{186} task filters out all the methods that may relate to the bug according to the names and comments of the covered methods.
(2) To address the issue of an excessive number of covered classes, we draw inspiration from an observation in test coverage. Specifically, we have observed that the fixing behavior tends to occur more frequently in classes that exhibit higher method-level coverage (for further details, please refer to Section \ref{subsec:imp}).
Based on this insight, we effectively mitigate the problem by selectively reducing the number of covered classes while ensuring that the performance remains unaffected, which is achieved by retaining only the Top-N classes that possess relatively high method-level coverage.
(3) For the documentation inconsistency, our insight is that as the LLMs have demonstrated powerful capabilities in code summarization~\cite{pu2023summarization,yang2023enhancing}, the {\em Source Code Reviewer} agent can be hired to fix the problematic documentation.
Therefore, we further design the \ding{185} {\em Method Doc Enhancement} task to handle the above problem.
Before finding related methods, the source code and the original comments of all covered methods in the suspicious class are listed in the prompt, the agent is then required to analyze the method call relationship to generate a new comment for each method.
The enhanced documentation can be seen from the bottom side of Figure \ref{fig:doc_enhance}.
In the first row, the task fills in the missing comment. 
In the second row, the task declares the callee method \texttt{process()} to alleviate the ``function nested'' problem, which provides more accurate information for the upcoming {\em Find Related Methods} task.

\subsection{Fault Confirmation}
\label{subsec:confirm}

{\bf Aim.}
In the previous steps, \toolname has only utilized the documentation and coverage information to search for the related methods in the codebase, which may not be enough to validate the exact fault location.
Therefore, we design the {\em Fault Confirmation} step to aggregate all useful information and make a final decision on which method is buggy.
Specifically, we build the \ding{187} {\em Method Review} task, where the {\em Software Test Engineer} agent examines the source code of the related methods and integrates all useful information to recognize the buggy methods.

{\bf Challenge.}
In order to review all related methods, a straightforward approach is to fill all the source codes in a single request and require the model to select the most suspicious method.
Nevertheless, we found in practice that as the volume of code in the context increases, it is more difficult for the model to focus on the fault location, resulting in a significant performance decrease.
Therefore, a strategy should be taken to validate all of the suspicious methods while also ensuring the model's accuracy.

{\bf Strategy.}
To address the above problem, we adopt a {\bf Multi-Round Dialogue} strategy in task \ding{187}, enabling the model to analyze one related method at a time.
The prompt template of the {\em Method Review} task is illustrated below:

\find{
One or more tests in the test class [TEST CLASS] failed:

Failed tests: [FAILED TESTS]


The method [METHOD NAME] may be problematic.


Detailed information is listed below:

[TEST INFOS]

Possible Causes: [POSSIBLE CAUSES],

Class of the Suspicious Method: [CLASS NAME]

Documentation of the Class: [CLASS DOC]

Suspicious Method Full Name: [METHOD NAME]

Suspicious Method Comment: [METHOD DOC]

Suspicious Method Code: [METHOD CODE]


As a Software Test Engineer, please carefully examine the code of the method [METHOD NAME] and determine if this method is the buggy location. You can return TRUE with the reason or only FALSE.

}

After the task \ding{187}, multiple methods can be regarded as suspicious.
Compared to spectrum-based and learning-based FL, \toolname does not generate suspicious values for ranking the suspicious methods.
For usability and evaluation purposes, we further design a \ding{188} {\em Get Top-1 Method} task in the last, where the {\em Software Test Engineer} agent 
retrospectively analyzes the buggy methods in all failed test classes to determine the most probable fault location.
In particular, if there is only one buggy method, the task takes it directly as the top-1 method.

\section{Experiment Design}
\label{sec:exp_set}

\subsection{Research Questions}
To assess the effectiveness of \toolname, we propose to answer the following research questions:
\begin{itemize}[leftmargin=*]
    \item {\bf RQ1:} {\bf What is the performance of \bftoolname in method-level fault localization?}
    This RQ aims to build a new effectiveness baseline for LLM-based method-level fault localization.
    
    \item {\bf RQ2:} {\bf How does our design choices affect the performance of \bftoolname?}
    This RQ could help measure the contribution of different components in \toolname, and thus inspire future studies.
    
    \item {\bf RQ3:} {\bf Can \bftoolname help developers in practice?} 
    This RQ conducts a user study to help understand how useful \toolname could be in practice.
    
\end{itemize}

\subsection{Benchmark}
\label{sec:benchmark}
In line with most existing fault localization work, we include the two versions of the widely-used benchmarks Defects4J~\cite{just2014defects4j} (i.e., Defects4J-V1.2.0 and Defects4J-V2.0.0) for evaluation. In particular, Defects4J-V1.2.0 consists of 395 bugs from 6 real-world Java projects (as shown in Table \ref{tab:benchmark}), on which we compare \toolname against all the studied baselines. In addition, Defects4J-V2.0.0 consists of additional 226 bugs from extra 9 Java projects, on which we further evaluate the generalization capability of studied techniques in the cross-project prediction setting in line with \grace~\cite{lou2021grace}. 


\subsection{Baselines}

{\bf LLM-based Baselines.}
We include two existing LLM-based FL techniques (\ie \llmao~\cite{yang2024llmao} and Wu \etal~\cite{wu2023llmfl}) as baselines. However, both techniques are only originally applied for fault localization within a small context, \ie \llmao~\cite{yang2024llmao} trains lightweight adapters on top of CodeGen~\cite{nijkamp2023codegen} to produce a suspicious score of each line within the given 128 code lines, and Wu \etal~\cite{wu2023llmfl} require ChatGPT to analyze the buggy line in a method with well-crafted prompts. Therefore, to adapt these techniques to project-level fault localization (\ie localizing the buggy method from the whole code base), we construct two variants of them (\ie \llmaoOchiai and \chatgptOchiai), which first leverage the state-of-the-art SBFL technique Ochiai~\cite{abreu2006ochiai} to reduce the context size and then apply these techniques for fault localization. In particular, for \chatgptOchiai, we apply a minimal change to the approach of Wu \etal by retaining their prompt format except that the command is changed from finding buggy lines to judging the correctness of the method, and ChatGPT is prompted to validate the top 20 suspicious methods localized by Ochiai. 
For \llmaoOchiai, given the top 20 suspicious methods localized by Ochiai, we first use \llmao to predict a suspicious value for each line in a method, then each method is re-ranked according to the value of its most suspicious line.

{\bf Statistic-based Baselines.}
We include the SBFL technique Ochiai~\cite{abreu2006ochiai}, FLUCCS~\cite{sohn2017fluccs} based on machine learning, DeepFL~\cite{li2019deepfl} based on deep learning, and \grace~\cite{lou2021grace} based on graph neural network.
For a fair comparison, we follow \grace~\cite{lou2021grace} to exclude mutation-related features in DeepFL and keep the features related to source code and coverage, the modified version is named DeepFL$_{cov}$.

\subsection{Implementation}
\label{subsec:imp}
We implement \toolname in Python based on the state-of-the-art multi-agent framework ChatDev~\cite{qian2023chatdev}, we use the {\em gpt-3.5-turbo-16k-0613} model of the ChatGPT family as the engine of \toolname.
For dynamic program analysis, we implement a lightweight JVM agent by the {\em java.lang.instrument}~\cite{instrument} package, and modify the framework of Defects4J to support dynamic program instrumentation.
To keep the program analysis focused on test/source code, we allow the tool to instrument the Java bytecode files in the specified folder.
For static program analysis, we use tree-sitter~\cite{treesitter} for code parsing and program element extraction.

We set the following alterable parameters to regulate the behavior of \toolname.
Among test information, the maximum token length of the test output log is 200.
The token length of class and method documentation is limited to 100.
To deal with the rare circumstances of overwhelming test cases, we keep no more than 5 failed test cases for each test class, which barely impairs the performance of \toolname in practice.

\begin{figure}[!t]
    \centering
    \includegraphics[width=\linewidth]{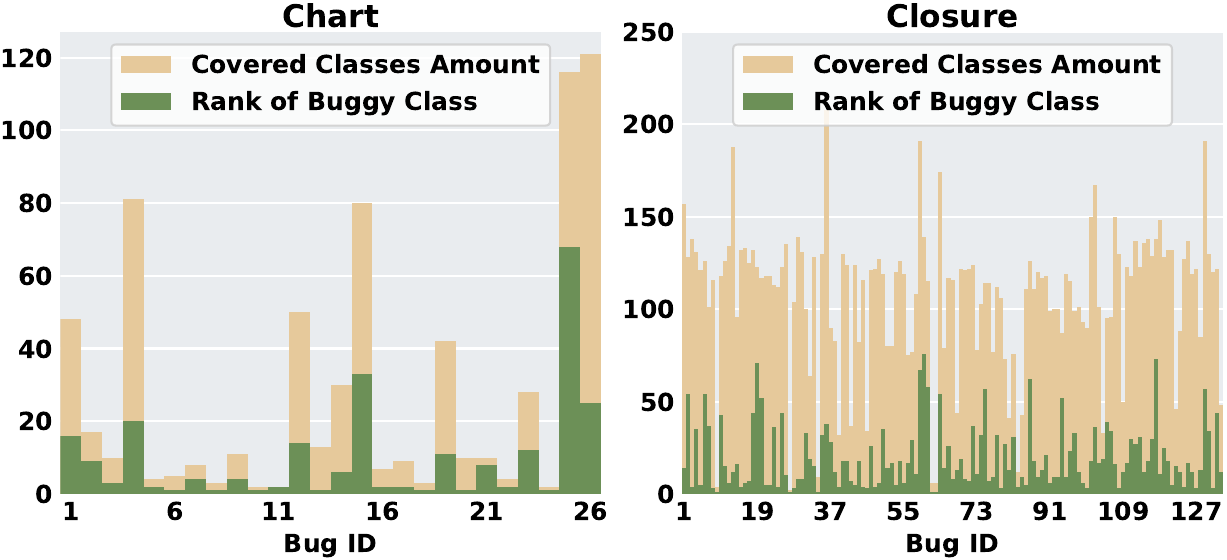}
    \caption{Method level coverage rank. For each bug, the green bar is the average coverage rank of all changed classes, and the orange bar is the amount of all covered classes. We show 2 out of 6 projects in Defects4J-V1.2.0 due to space limit.}
    \label{fig:classes}
\end{figure}

For covered class reduction (introduced in Section \ref{subsec:navi}), we investigate the method-level coverage rank of the class changed by developers among all covered classes.
Specifically, the method-level coverage rate of class $c$ is calculated with $r_{c} = (\sum_{i=1}^{N} \mathbb{I}_{y_{i}}) / N$, where $\mathbb{I}_{y_{i}}$ is an function to indicate whether method $y_{i}$ is covered.
Based on the results shown in Figure \ref{fig:classes}, we retain classes that fall in the top-50 method-level coverage rates, which ensures the buggy class is preserved in over 98\% of the cases.

\begin{table}[!t]
\centering
\caption{Information of Defects4J-V1.2.0 benchmark.}
\label{tab:benchmark}
\begin{tabular}{llrr}
\hline
\textbf{ID} & \textbf{Name}           & \textbf{\#Bug} & \textbf{LOC(k)} \\ \hline
Chart       & JFreeChart              & 26             & 96              \\
\rowcolor{grey} Lang        & Apache commons-lang     & 65             & 22              \\
Math        & Apache commons-math     & 106            & 85              \\
\rowcolor{grey} Time        & Joda-Time               & 27             & 28              \\
Mockito     & Mockito framework       & 38             & 23              \\
\rowcolor{grey} Closure     & Google Closure compiler & 133            & 90              \\ \hline
\end{tabular}
\end{table}

\subsection{Metrics}
\label{subsec:metrics}
Following prior studies~\cite{sohn2017fluccs, li2019deepfl, lou2021grace, qian2023gnet4fl}, we assess the effectiveness of \toolname based on the Top-N (N=1,3,5) metric.
Top-N computes the number of bugs that have at least one buggy element localized within the first N positions in the ranked list.
It is easy to calculate the Top-1 value with the design of {\em Get Top-1 Method} task in \toolname.
However, since the ChatGPT model is a master at judging true or false rather than calculating suspicious values, it is non-trivial to get Top-3 and Top-5.
To alleviate the problem, we compromise by keeping a list of all the methods that \toolname considers suspicious in chronological order and calculating Top-N with the rank of the buggy methods in the list.

\section{Evaluation}
\label{sec:eval}

\subsection{RQ1: Localization Performance}
\label{subsec:rq1}
{\bf Compared to LLM-based baselines.}
We first compare \toolname with \chatgptOchiai and \llmaoOchiai that both use LLMs to drive fault localization, the results are shown in Table \ref{tab:llm_d4j140}.
Overall, \toolname achieves reasonable performance: it can localize 157 out of 395 bugs within Top-1, 28 more than \llmaoOchiai and 36 more than \chatgptOchiai.
When considering the Top-1 metric on the individual project, we observe that \toolname significantly outperforms the other two techniques on nearly every project except Closure, where \llmaoOchiai localizes 18 more methods within Top-1.
In terms of Top-3 and Top-5 metrics, we found that although \toolname outperforms \chatgptOchiai, it failed to localize more bugs than \llmaoOchiai.

The above results can be explained by the different mechanisms of \llmaoOchiai and \toolname.
The \llmaoOchiai predicts a suspicious value for each method that has been pre-localized by Ochiai, which is more talented at ranking tasks than the ChatGPT model.
However, since \toolname behaves more like a human by browsing the entire codebase without any auxiliary localization results, the limited search space and the ranking issue~\cite{qin2023rank} of the LLMs hinder it from achieving better performance in Top-3 and Top-5 metrics.

\begin{table}[!t]
\centering
\caption{Comparision with LLM-based baselines on Defects4J-V1.2.0.}
\label{tab:llm_d4j140}
 \resizebox{0.9\linewidth}{!}
{
\begin{tabular}{l|c|lrrr}
\hline
\textbf{Project} & \textbf{\# Bugs} & \textbf{Techniques} & \textbf{Top1} & \textbf{Top3} & \textbf{Top5} \\ \hline
\multirow{3}{*}{Chart}  & \multirow{3}{*}{26} & ChatGPT$_{Ochiai}$ & 11    & 15    & 15    \\
 & & LLMAO$_{Ochiai}$ & 11 & 15 & 19     \\ \cline{3-6} 
& & \cellcolor{grey} \toolname       & \cellcolor{grey} 16    & \cellcolor{grey} 18    & \cellcolor{grey} 19    \\ \hline
\multirow{3}{*}{Lang}   & \multirow{3}{*}{65} & ChatGPT$_{Ochiai}$ & 29    & 35    & 36    \\
 & & LLMAO$_{Ochiai}$   & 31 & 49 & 54     \\ \cline{3-6} 
& & \cellcolor{grey} \toolname       & \cellcolor{grey} 44    & \cellcolor{grey} 45    & \cellcolor{grey} 45    \\ \hline
\multirow{3}{*}{Math}   & \multirow{3}{*}{106}    & ChatGPT$_{Ochiai}$ & 38    & 55    & 55    \\
 & & LLMAO$_{Ochiai}$   & 30 & 59 & 70     \\ \cline{3-6} 
& & \cellcolor{grey} \toolname       & \cellcolor{grey} 49    & \cellcolor{grey} 60    & \cellcolor{grey} 61    \\ \hline
\multirow{3}{*}{Time}   & \multirow{3}{*}{27} & ChatGPT$_{Ochiai}$ & 9 & 11    & 12    \\
 & & LLMAO$_{Ochiai}$   & 7 & 13 & 14     \\ \cline{3-6} 
& & \cellcolor{grey} \toolname       & \cellcolor{grey} 11    & \cellcolor{grey} 13    & \cellcolor{grey} 13    \\ \hline
\multirow{3}{*}{Mockito}    & \multirow{3}{*}{38} & ChatGPT$_{Ochiai}$ & 12    & 16    & 16    \\
 & & LLMAO$_{Ochiai}$   & 8 & 19 & 26     \\ \cline{3-6} 
& & \cellcolor{grey} \toolname       & \cellcolor{grey} 13    & \cellcolor{grey} 14    & \cellcolor{grey} 14    \\ \hline
\multirow{3}{*}{Closure}    & \multirow{3}{*}{133}    & ChatGPT$_{Ochiai}$ & 22    & 36    & 38    \\
 & & LLMAO$_{Ochiai}$   & 42 & 57 & 70     \\ \cline{3-6} 
& & \cellcolor{grey} \toolname       & \cellcolor{grey} 24    & \cellcolor{grey} 33    & \cellcolor{grey} 35    \\ \hline \hline
\multirow{3}{*}{Overall}    & \multirow{3}{*}{395}    & ChatGPT$_{Ochiai}$ & 121   & 168   & 172   \\
 & & LLMAO$_{Ochiai}$   & 129 & 212 & 253     \\ \cline{3-6} 
& & \cellcolor{grey} \toolname       & \cellcolor{grey} 157    & \cellcolor{grey} 183    & \cellcolor{grey} 187    \\ \hline
\end{tabular}
}
\end{table}

{\bf Compared to statistic-based baselines.}
To investigate how \toolname performs against the statistic-based techniques, we select four techniques based on different methodologies: FLUCCS~\cite{sohn2017fluccs} (machine learning), DeepFL$_{cov}$~\cite{li2019deepfl} (multi-layer perception), \grace~\cite{lou2021grace} (graph neural network), and Ochiai~\cite{abreu2006ochiai} (suspicious formula).
From Table \ref{tab:stt_d4j140}, we observe that as a first attempt to use the LLMs for method-level fault localization, \toolname is not yet able to surpass existing approaches entirely.
Overall, \toolname localizes 157 of 395 bugs within Top-1, 77 more than Ochiai, 3 less than FLUCCS, 19 less than DeepFL$_{cov}$, and 35 less than \grace.

However, when focusing on the results of individual projects, we find that \toolname has achieved competitive performance in all software projects except Closure.
To explain this, we scrutinized every bug that \toolname failed to localize while other approaches were able to make it.
We attribute the prominent weak effectiveness of \toolname on Closure to three aspects: 
(1) Some bugs in Closure possess very similar characteristics.
For instance, bugs Closure-18 and Closure-31 have failed test cases with similar purposes (\texttt{testDependencySorting} and \texttt{testDependencySortingWhitespaceMode}), and even the same buggy method \texttt{parseInputs()}.
In these cases, the learning-based techniques learn much richer in-project knowledge than \toolname as they perform within-project prediction with the leave-one-out training strategy. \toolname, however, needs to localize a bug from scratch based on more generalized knowledge from ChatGPT.
(2) The tests in Closure tend to cover more program elements.
For example, the average number of classes covered by the bugs in Closure is 107.7, which is extremely larger than those in Lang (2.3) and Chart (27.5).
In this case, it is more difficult for \toolname to recognize the suspicious class at once among a large number of covered classes.
(3) Since \toolname is designed to validate rather than rank the methods, the number of suspicious methods in the result of \toolname is relatively small.
Among all 395 bugs, there are 319 (374) cases where the number of suspicious methods in the result is less than 3 (5).
Consequently, the increase in the number of bugs localized by \toolname becomes progressively smaller from Top-1 to Top-3 (157 to 183) and then Top-3 to Top-5 (183 to 187).

\begin{table*}[!t]
\centering
\caption{Comparision with statistic-based baselines on Defects4J-V1.2.0.}
\label{tab:stt_d4j140}
 \resizebox{0.95\linewidth}{!}
{
\begin{tabular}{lr|rrr|rrr|rrr|rrr|rrr}
\hline
\multicolumn{1}{c}{\multirow{2}{*}{\textbf{Project}}} & \multirow{2}{*}{\textbf{\# Bugs}} & \multicolumn{3}{c|}{\bftoolname} & \multicolumn{3}{c|}{\textbf{GRACE}} & \multicolumn{3}{c|}{\textbf{DeepFL}$_{cov}$} & \multicolumn{3}{c|}{\textbf{FLUCCS}} & \multicolumn{3}{c}{\textbf{Ochiai}} \\ \cline{3-17} 
 & & \textbf{Top1} & \textbf{Top3} & \textbf{Top5} & \textbf{Top1} & \textbf{Top3} & \textbf{Top5} & \textbf{Top1} & \textbf{Top3} & \textbf{Top5} & \textbf{Top1} & \textbf{Top3} & \textbf{Top5} & \textbf{Top1} & \textbf{Top3} & \textbf{Top5} \\ \hline
Chart & 26 & 16 & 18 & 19  & 14 & 20 & 22  & 12 & 18 & 21  & 15 & 19 & 16  & 6  & 14 & 15 \\
\rowcolor{grey} Lang  & 65 & 44 & 45 & 45  & 42 & 54 & 57  & 43 & 53 & 56  & 40 & 53 & 55  & 24 & 44 & 50 \\
Math  & 106   & 49 & 60 & 61  & 61 & 78 & 89  & 39 & 68 & 80  & 48 & 77 & 83  & 23 & 52 & 62 \\
\rowcolor{grey} Time  & 27 & 11 & 13 & 13  & 11 & 14 & 19  & 9 & 16 & 18  & 8  & 15 & 18  & 6  & 11 & 13 \\
Mockito               & 38 & 13 & 14 & 14  & 17 & 24 & 26  & 9 & 15 & 21  & 7  & 19 & 22  & 7  & 14 & 18 \\
\rowcolor{grey} Closure               & 133   & 24 & 33 & 35  & 47 & 70 & 81  & 64 & 86 & 97  & 42 & 66 & 77  & 14 & 30 & 38 \\ \hline \hline
Overall               & 395   & 157    & 183    & 187 & 192    & 260    & 294 & 176    & 256    & 293 & 160    & 249    & 271 & 80 & 165    & 196    \\ \hline
\rowcolor{grey} w/o Closure    & 262   & 133    & 150    & 152 & 145    & 190    & 213 & 112    & 170    & 196 & 118    & 183    & 194 & 66 & 135    & 158    \\ \hline
\end{tabular}
}
\end{table*}

The last row of Table \ref{tab:stt_d4j140} presents the result without project Closure, \toolname reaches a similar level as \grace in terms of Top-1 and outperforms the other three techniques on each project by localizing 67 more bugs than Ochiai, 15 more than FLUCCS, 21 more than DeepFL$_{cov}$ and only 12 less than \grace.
For cross-project evaluation, we follow Lou \etal~\cite{lou2021grace} to evaluate \toolname against other statistic-based approaches on 226 extra bugs in Defects4J-V2.0.0, the result in Table \ref{tab:res_d4j200} shows that \toolname significantly outperforms most of the other approaches by localizing 35 more bugs within Top-1 than DeepFL$_{cov}$, 21 more than FLUCCS, and 46 more than Ochiai.
\toolname gains comparable performance to \grace with only 7 fewer bugs localized in the top.

\begin{figure}[!t]
    \centering
    \includegraphics[width=0.5\linewidth]{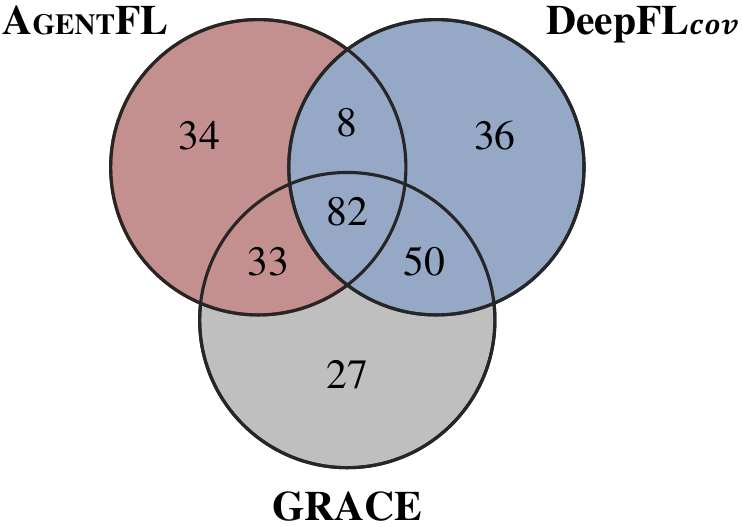}
    \caption{Overlap results on Defects4J-V1.2.0.}
    \label{fig:venny}
\end{figure}

We also investigated the overlaps among the bugs localized within Top-1 by different approaches on the Defects4J-V1.2.0 benchmark, and the results are shown in Figure \ref{fig:venny}.
The figure illustrates the complementarity of \toolname with existing approaches on localizing different bugs. Specifically, 34 bugs can be uniquely localized within Top-1 by \toolname while the numbers for DeepFL$_{cov}$ and \grace are 36 and 27 respectively. 

\begin{table}[!t]
\centering
\caption{Results on Defects4J-V2.0.0.}
\label{tab:res_d4j200}
 \resizebox{0.9\linewidth}{!}
{
\begin{tabular}{c|c|lrrr}
\hline
\textbf{Project} & \textbf{\# Bugs} & \textbf{Techniques}  & \textbf{Top1} & \textbf{Top3} & \textbf{Top5} \\ \hline
\multirow{5}{*}{Overall} & \multirow{5}{*}{226}    & Ochiai      & 32   & 74   & 93   \\
& & FLUCCS      & 57   & 97   & 119  \\
& & DeepFL$_{cov}$ & 43   & 89   & 112  \\
& & GRACE       & 85   & 119  & 140  \\ \cline{3-6} 
& & \cellcolor{grey} \toolname    & \cellcolor{grey} 78   & \cellcolor{grey} 98   & \cellcolor{grey} 103  \\ \hline
\end{tabular}
}
\end{table}

\notez{
{\bf Answer to RQ1:}
\toolname can effectively localize the bugs in Defects4J, which outperforms the other LLM-based approaches and shows complementarity with existing statistic-based techniques in terms of the Top-1 metric.
}

\subsection{RQ2: Ablation Study}
\label{subsec:atten}

For this research question, we aim to investigate the contributions of different design choices to \toolname.
To this end, we separately remove the {\em Test Behavior Analysis} task, {\em Test Failure Analysis} task, and {\em Method Doc Enhancement} task from \toolname, and retain the other tasks that need to maintain the proper functionality of the system.
Note that after removing the Test Failure Analysis task, the Test Behavior Analysis task is also muted, since the results of the former task participate in the construction of the latter task's prompt.

The performances of the variants of \toolname are shown in Table \ref{tab:ablation}.
We notice that when the {\em Test Failure Analysis} task is discarded, the efficacy of \toolname decreases sharply, with the number of bugs localized within Top-1 decreasing from 157 to 125.
This indicates that it is a finer choice to offer ChatGPT time to think about the root causes of the test failure and take the analysis result as intermediate data to guide the subsequent localization process.
This pattern is more in line with the debugging behavior of developers and shares the idea of chain-of-thought~\cite{wei2022cot}.
Similarly, when the {\em Method Doc Enhancement} task is removed, the Top-1 result drops significantly from 157 to 133.
While this phenomenon emphasizes the importance of higher quality documentation for \toolname to more accurately localize the buggy methods, it also calls on the developers to write better documentation during production, as the human wisdom preserved in documents can conversely support the development and maintenance of software.

Surprisingly, when we abandon the {\em Test Behavior Analysis} task, the performance of \toolname declines marginally from 157 to 148 on the Top-1 metric.
To deeply understand the reasons for this phenomenon, we analyze the characteristics of utility methods corresponding to failed test cases.
We find that although each test method calls 4.93 other utility methods on average, these utility methods are generally shared by multiple test methods to handle common operations such as test initialization.
This means that the code of the test method may already carry most of the information needed by the ChatGPT to understand the behavior of the test.
Nonetheless, the fact of performance gain still emphasizes the potential of other sources of information than just the test code.

\notez{
{\bf Answer to RQ2:}
The Test Behavior Analysis, Test Failure Analysis, and Method Doc Enhancement tasks are useful to underpin the effectiveness of \toolname. The contribution of Test Behavior Analysis appears to be relatively small.
}

\subsection{RQ3: User study}
In this RQ, we intend to evaluate to what extent can \toolname help developers localize bugs in practice.
Kochhar \etal~\cite{kochhar2016practitioner} found that 85\% of the practitioners strongly agreed that the ability to provide rationale is important, with one of the respondents stating {\em ``because to make a decision about bug ﬁxing I want to exactly know why the automated tool `thinks' that the code has a bug''}.
Therefore, merely relying on the Top-N metric may not be enough to reflect the usability of \toolname,
since it not only lists the suspicious methods but also provides the user with additional natural language interpretations.
To investigate the usability of \toolname, we re-evaluate \toolname through the following user study:
(1) {\em Participant Selection}.
We recruited five Ph.D. students each of whom has more than five years of programming experience and does not know about the Defects4J benchmark before;
(2) {\em Data Preparation}.
By excluding 98 out of 395 bugs where \toolname did not find any suspicious methods, we constructed 297 bug-suggestion pairs. The suggestion for each bug contains the top-3 suspicious methods and the corresponding rationales generated by \toolname;
(3) {\em Localization Procedure}.
We ask each participant to localize the buggy method from the codebase with the test failure information (including failed tests, error stack traces, and test outputs) and the suggestions provided by \toolname.
The participants are allowed to use web searches if they are not familiar with any program logic.
Given that the previous study shows that a practitioner usually takes about 11 minutes to localize a bug with an FL tool~\cite{xia2016userstudy}, we set the time limit to 15 minutes, providing the users with sufficient time to finish the task;
(4) {\em Results Collection}.
Finally, for each bug-suggestion pair, we regard \toolname as helpful if the user successfully localizes the buggy method within the specified time.

\begin{table}[!t]
\centering
\caption{Result of ablation study.}
\label{tab:ablation}
 \resizebox{\linewidth}{!}
{
\begin{tabular}{c|c|lrrr}
\hline
\textbf{Project} & \textbf{\# Bugs} & \textbf{Techniques}               & \textbf{Top1} & \textbf{Top3} & \textbf{Top5} \\ \hline
\multirow{4}{*}{Overall} & \multirow{4}{*}{395} & w/o TestBehaviorAnalysis             & 148  & 174  & 175  \\
 & & w/o TestFailureAnalysis  & 125  & 148  & 152  \\
 & & w/o MethodDocEnhancement & 133  & 156  & 160  \\ \cline{3-6} 
 & & \cellcolor{grey} \toolname & \cellcolor{grey} 157  & \cellcolor{grey} 183  & \cellcolor{grey} 187  \\ \hline
\end{tabular}
}
\end{table}

\begin{figure}[!t]
    \centering
    \includegraphics[width=\linewidth]{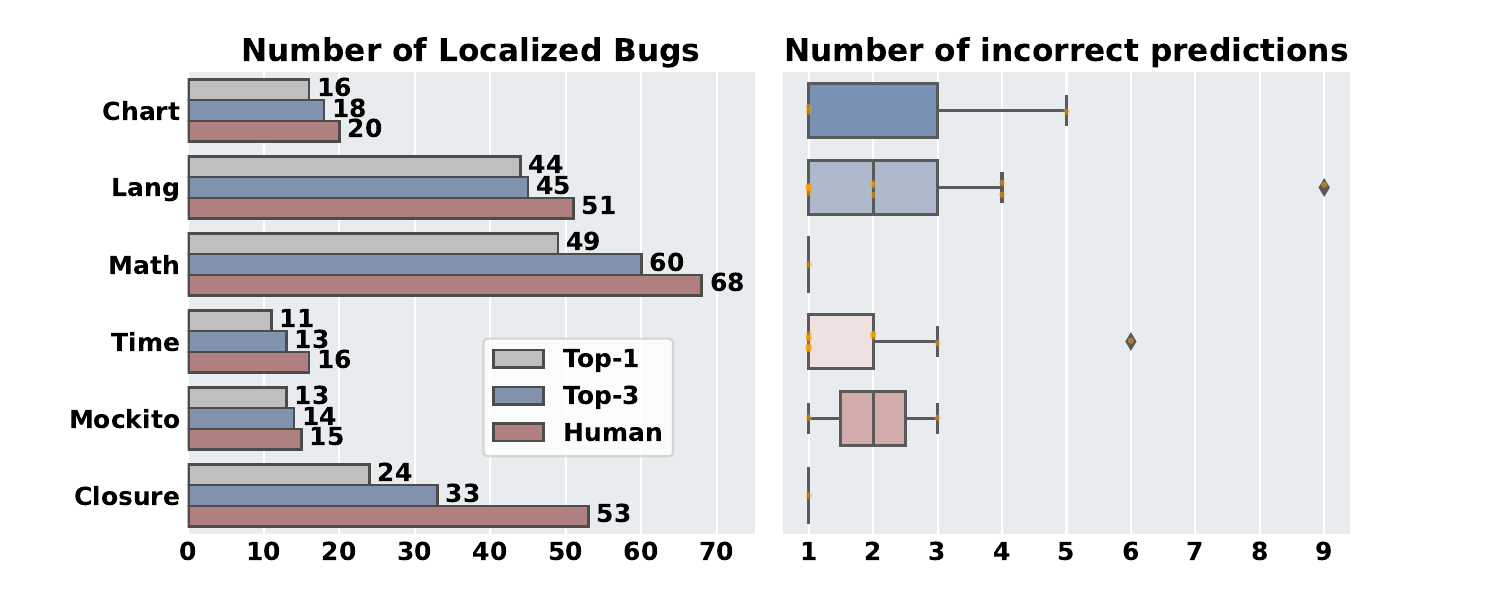}
    \caption{Evaluation results of RQ3.}
    \label{fig:rq3}
\end{figure}

The result of our user study is shown in Figure \ref{fig:rq3}, from which we observe that with the help of \toolname, developers can consistently localize more bugs in all the projects compared with the vanilla \toolname. 
Two most significant breakthroughs occur in projects Math and Closure, where the numbers of localized bugs in terms of the Top-1 (Top-3) results are improved by 19 (8) and 29 (20), respectively.
To understand why participants achieved better localization results, we further investigate the cases where \toolname failed to suggest the buggy method at the top position.
The statistics show that in more than 80\% of the cases, the number of incorrect predictions (\ie non-buggy methods that are ranked before buggy methods) produced by \toolname does not exceed 3, thus bringing less distraction to the users.
Such results indicate that the explanations produced by \toolname could provide additional fault-related information for the developers and help them better finish the fault localization task, demonstrating the potential usability of \toolname in practice.

\begin{figure}[!t]
    \centering
    \includegraphics[width=0.9\linewidth]{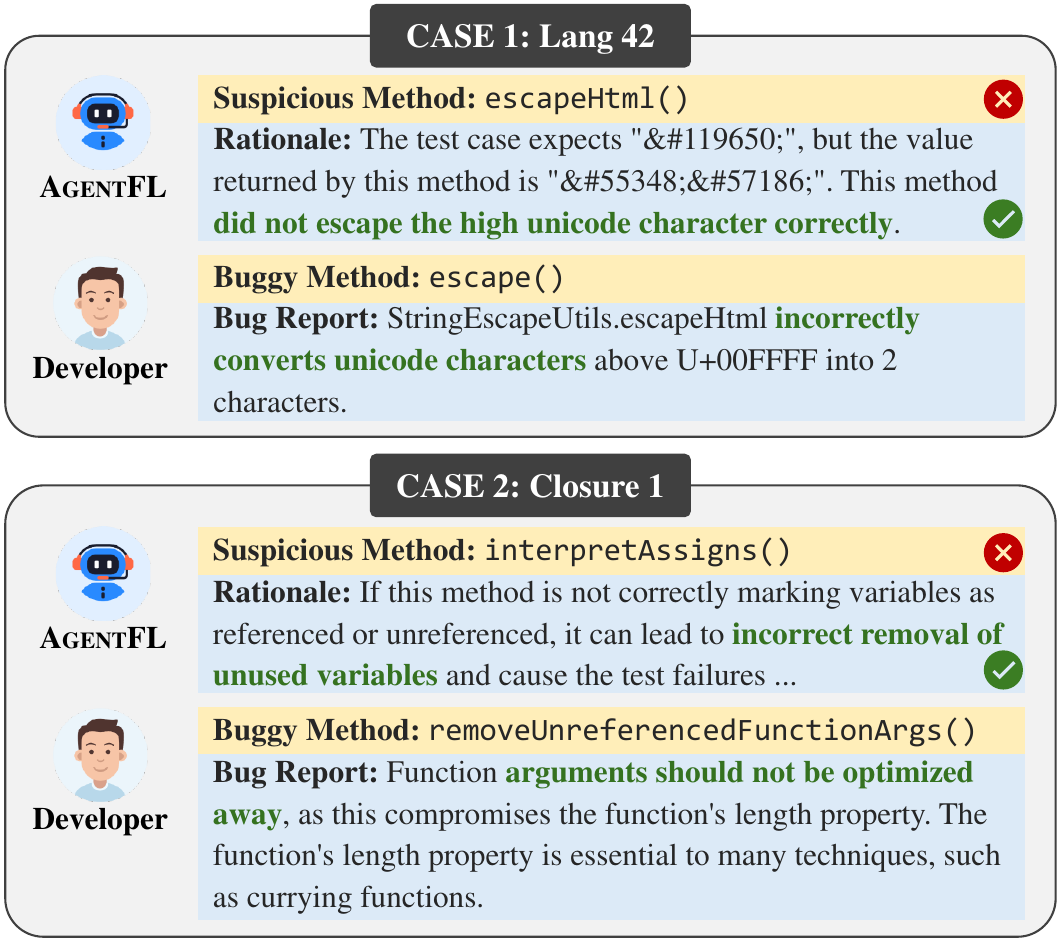}
    \caption{Two use cases for bug Lang-42 and Closure-1.}
    \label{fig:case}
\end{figure}

To help further understand the usability of \toolname in practice, we demonstrate two cases in Figure \ref{fig:case}. 
In the first case, \toolname successfully understands the root cause of the bug Lang-42 (\ie ``did not escape the high unicode character correctly'').
However, although the method \texttt{escapeHtml} has been mentioned in the bug report, the developer eventually modified its callee method \texttt{escape} (\ie the method called by \texttt{escapeHtml} to handle with more specific character escaping logic) to fix the bug.
Similarly, in the second case, \toolname still failed to localize the buggy method, but it correctly interpreted the root cause of the bug Closure-1 and suggested the users to investigate the method \texttt{interpretAssigns}, which is exactly the caller of the buggy method. As shown in the figure, the incorrect removal of unused variables happens in the method \texttt{removeUnreferencedFunctionArgs}, making it the buggy location.
For both cases, \toolname provides reasonable explanations despite its prediction (\ie the caller methods of the buggy methods) is not exactly precise, which allows the user to {\em ``find the error location with a little extra work''} (one participant left such a comment).



\notez{
{\bf Answer to RQ3:}
\toolname exhibits relatively high usability in practice with its few incorrect predictions and the provision of rationale.
}
\section{Discussion}
\label{sec:dis}

\subsection{Cost Analysis}
\label{subsec:cost}

\begin{figure}[!t]
    \centering
    \includegraphics[width=\linewidth]{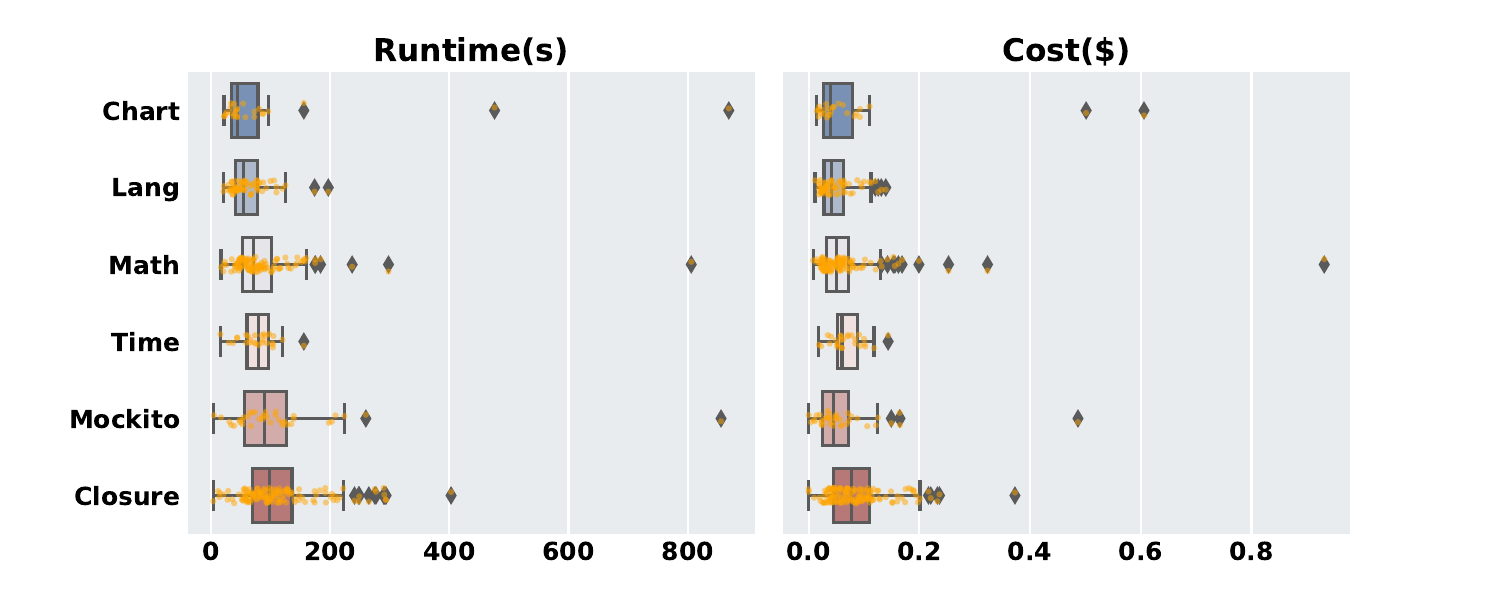}
    \caption{Cost analysis results.}
    \label{fig:cost}
\end{figure}

We assessed the performance of \toolname on the often-concerned issues of prices and time consumption.
The results are demonstrated in Figure \ref{fig:cost}, where each sample is marked with an orange dot.
For runtime, we record the seconds required for \toolname to completely deal with each bug, including test runs, program analysis, and ChatGPT inference.
For cost, we calculate how many dollars \toolname needs to localize each bug by multiplying the total number of consumed tokens with the model price (0.003 dollar per thousand tokens).
We observe that \toolname offers affordable cost in terms of time and money, which only takes an average of 0.074 dollars and 97 seconds to localize a fault.
In over 95\% cases, the cost per bug would not exceed 0.2 dollars and 200 seconds.
Such a low expense is expected to mitigate the expensive costs that developers spend on FL.

\subsection{Threats to Validity}
{\bf Internal.}
The main internal threat comes from the data leakage problem.
As the training data of the {\em gpt-3.5-turbo-16k-0613} model is up to Sep 2021 while Defects4J-V1.2.0 is released in Feb 2018, the fault localization dataset may have been used for model training.
We mitigate this threat by making sure that the input to ChatGPT does not include any content related to the project name, human-written bug report, or bug ID.
Moreover, the poor performance of the default ChatGPT (\ie the baseline ChatGPT$_{Ochiai}$) also indicates that ChatGPT has not simply memorized the answer, while the significant improvement of \toolname (compared to the default ChatGPT) shows the effectiveness of our approach.

{\bf External.}
The main external threat to validity comes from our evaluation benchmark.
The results obtained by \toolname may not generalize to other benchmarks.
To address this, we evaluate \toolname not only on the Defects4J-V1.2.0 benchmark but also on Defects4J-V2.0.0 to demonstrate the generalizability.

\section{Conclusion}
\label{sec:conc}
In this paper, we introduce \toolname, a multi-agent system based on ChatGPT for automated LLM-based FL.
Inspired by the human debugging actions, \toolname breaks the localization process into three steps (Fault Comprehension, Codebase Navigation, and Fault Confirmation), and hires several agents in each step to handle specific tasks.
The evaluation results on the Defects4J-V1.2.0 benchmark show that \toolname outperforms other LLM-based approaches and can be used complementarily to the existing learning-based techniques.
A user study also demonstrates the potential of \toolname in helping developers localize real-world software bugs.
Finally, the cost analysis shows that \toolname takes only an average of 0.074 dollars and 97 seconds to localize 157 out of 395 bugs within Top-1.






\end{document}